%% file: JSSBiDAG.march2020.tex
\documentclass[article]{jss}
\usepackage{caption}
\usepackage{natbib}
\usepackage{subcaption}
\usepackage{amsmath}
\usepackage[linesnumbered,ruled,vlined]{algorithm2e}

\newcommand{\imc}{\code{itera\-tive\-MCMC} }

\newcommand{\omc}{\code{or\-der\-MCMC} }

\newcommand{\imcc}{\code{itera\-tive\-MCMC}}

\newcommand{\omcc}{\code{or\-der\-MCMC}}

\author{Polina Suter\\ETH Z\"urich 
   \And Jack Kuipers\\ETH Z\"urich
   \And Giusi Moffa\\University of Basel
   \And Niko Beerenwinkel\\ETH Z\"urich}
\Plainauthor{Polina Suter, Jack Kuipers, Giusi Moffa, Niko Beerenwinkel}

\title{Bayesian structure learning and sampling of Bayesian networks with the \proglang{R} package \pkg{BiDAG}}
\Plaintitle{Bayesian structure learning and sampling of Bayesian networks with the R package BiDAG}
\Shorttitle{Bayesian structure learning and sampling with \pkg{BiDAG}}

\Abstract{
The R package \pkg{BiDAG} implements Markov chain Monte Carlo (MCMC) methods for structure learning and sampling of Bayesian networks. The package includes tools to search for a maximum {\em a posteriori} (MAP) graph and to sample graphs from the posterior distribution given the data. A new hybrid approach to structure learning enables inference in large graphs.  In the first step, we define a reduced search space by means of the PC algorithm or based on prior knowledge.  In the second step, an iterative order MCMC scheme proceeds to optimize the restricted search space and estimate the MAP graph. Sampling from the posterior distribution is implemented using either order or partition MCMC.  The models and algorithms can handle both discrete and continuous data. The \pkg{BiDAG} package also provides an implementation of MCMC schemes for structure learning and sampling of dynamic Bayesian networks. 
}
\Keywords{Bayesian networks, dynamic Bayesian networks, structure learning,  Bayesian inference, MCMC, \proglang{R}}
\Plainkeywords{Bayesian networks, dynamic Bayesian networks, directed acyclic graphs, DAGs, structure learning,  Bayesian inference, posterior DAG structure sampling, MCMC, R}
\Address{
  Polina Suter\\
  ETH Zurich \\
  Department of Biosystems Science and Engineering\\
  Mattenstrasse 26\\
  4058 Basel, Switzerland\\
  E-mail: \email{polina.suter@bsse.ethz.ch}\\
  \emph{and}\\
  SIB Swiss Institute of Bioinformatics\\
  4058 Basel, Switzerland \\

  Jack Kuipers\\
  ETH Zurich \\
  Department of Biosystems Science and Engineering\\
  Mattenstrasse 26\\
  4058 Basel, Switzerland\\
  E-mail: \email{jack.kuipers@bsse.ethz.ch}\\
  \emph{and}\\
  SIB Swiss Institute of Bioinformatics\\
  4058 Basel, Switzerland \\

  Giusi Moffa\\
  University of Basel \\
  Department of Mathematics and Computer Science\\
  Spiegelgasse 1 \\
  4051 Basel, Switzerland\\
  E-mail: \email{giusi.moffa@unibas.ch}\\

  Niko Beerenwinkel\\
  ETH Zurich \\
  Department of Biosystems Science and Engineering\\
  Mattenstrasse 26\\
  4058 Basel, Switzerland\\
  E-mail: \email{niko.beerenwinkel@bsse.ethz.ch}\\
  \emph{and}\\
  SIB Swiss Institute of Bioinformatics\\
  4058 Basel, Switzerland
}

\begin{document}

\input{chapters/Introduction}
\input{chapters/Background}
\input{chapters/Package}

\input{chapters/DBNs}

\input{chapters/Examples}
\input{chapters/Application}
\input{chapters/Runtime}

\input{chapters/Discussion}

\newpage  

\bibliography{myReferences}
\end{document}

%% file: chapters/Introduction.tex
\section{Introduction} 
A Bayesian network is a probabilistic graphical model, which represents conditional independence relationships between a set of random variables by a directed acyclic graph (DAG).The problem of DAG learning from observational data is hard \citep{npcompl}, and the number of DAGs grows super-exponentially with the number of nodes. Hence, developing and implementing methods to learn an underlying DAG from observational data in reasonable time continues to be the focus of much research \citep{Bartlett2017,Gibbs2016,rblip}. \cite{Drton2017} provide an overview of the approaches for structure learning of graphical models including Bayesian networks. 

The \proglang{R} \citep{Rcit} packages \pkg{pcalg} \citep{pcalg}, \pkg{BNlearn} \citep{bnlearn}, \pkg{bnstruct} \citep{bnstruct} and the \proglang{Java}-based toolbox \pkg{TETRAD} \citep{TETRAD} implement multiple approaches to structure learning, including both constraint-based and search-and-score methods. Constraint-based methods use conditional independence tests to learn the edges of the graph. Search-and-score methods rely on an efficient search strategy in the space of DAGs and a score function to find the graph which best explains the data. Hybrid methods, such as max-min hill climbing \citep{MMHC}, use a combination of both approaches to learn the optimal graph. A comparison of accuracy and efficiency of different methods for structure learning of Bayesian networks can be found in \cite{Scutari2019}. Despite a wide variety of available tools, most of them aim to find one best solution.  However, especially when the number of observations is modest, relying on one best graph can be inadequate because many graphs may explain the data similarly well \citep{bayes2003}. Bayesian methods can help to address this issue. Posterior sampling, in particular, enables Bayesian model averaging and allows us to adequately account for modeling uncertainty when the number of observations is finite. However, only a few tools exist for Bayesian structure learning, probably because Bayesian approaches are computationally demanding and unfeasible in large domains. The \proglang{R} package \pkg{mcmcabn} \citep{mcmcabn} implements a structure MCMC algorithm for sampling DAGs from their posterior distribution given the data.  Structure MCMC is only suitable for domains with a limited number of nodes. The \proglang{R} package \pkg{deal} implements Bayesian parameter learning, but for structure learning, it uses a greedy search with random restarts \citep{deal}. The \proglang{Matlab/C/Java} package \pkg{BDagl} \citep{bdagl} implements an order MCMC scheme \citep{bayes2003} without any restrictions on the search space, so that it is only feasible for small networks and does not scale well beyond 20 nodes. \par

Here, we describe the \proglang{R} package \pkg{BiDAG}, an implementation of various MCMC schemes, which overcomes the issues previously restricting Bayesian methods to small domains. \pkg{BiDAG} implements order \citep{bayes2003} and partition \citep{Partition} MCMC schemes. These scales to networks with hundreds of nodes when combined with the hybrid approach of \cite{plus1}. Both order and partition MCMC schemes can sample from the posterior and find a MAP DAG, and both reach convergence much faster than the structure MCMC approach.  Simulation studies have shown that the iterative order MCMC scheme \citep{plus1} displays better accuracy to discover the ground truth DAG compared to other well-established methods such as the PC algorithm \citep{PC2000} or greedy equivalent search (GES) \citep{npcompl}. \par

The \pkg{BiDAG} software supports both discrete and continuous data types, and the methods also apply to weighted data as required, for example, in mixture models \citep{Kuipers2018}. Further, all the implemented MCMC schemes handle structure learning and sampling of first-order dynamic Bayesian networks (DBNs).
\pkg{BiDAG} is available from the Comprehensive R Archive Network (CRAN) at http://CRAN.R-project.org/package=BiDAG.\par

\pkg{BiDAG} also imports some methods from other packages. In the first step of a hybrid approach, it uses by default the constraint-based PC algorithm from \pkg{pcalg} to define a search space, complemented by a new implementation of conditional independence tests for discrete and weighted data. \pkg{BiDAG} includes a visualization tool, which imports methods from the \proglang{R} packages  \pkg{graph} \citep{graph} and \pkg{Rgraphviz} \citep{rgraph}.\par

In Section \ref{methback}, we describe the methodological background behind the algorithms in \pkg{BiDAG}. In Section \ref{bidag}, we describe the \pkg{BiDAG} functions, further divided into four subsections on structure learning and sampling, posterior model selection, convergence diagnostics, and model comparison. In Section \ref{DBNs}, we describe how to use the package for structure learning of DBNs. Section \ref{examples} contains examples of structure learning and sampling on two simulated data sets. In this section we also show how sampling from the posterior can improve model selection. In Section \ref{sec:app}, we apply the package to the problem of characterizing cancer sub-types. Finally, in Section \ref{runtime}, we discuss the runtime of the implemented algorithms in different simulation settings. \\

%% file: chapters/Background.tex
\section{Methodological background}
\label{methback}
A Bayesian network describes a factorization of a joint probability distribution $P(\textbf{X})$ of a set of random variables $\textbf{X}=(X_{1},...,X_{n})$ by means of a DAG. Specifically we can define a Bayesian network $\mathcal{B}$ as a pair $(\mathcal{G}, \Theta)$ where $\mathcal{G}$ is a DAG whose nodes represent the random variables in $\textbf{X}$ and  $\Theta$  are the parameters of the probability distributions $P(X_{i} \mid \mathbf{Pa}_{i})$ describing the relationship between each variable $X_{i}$ and its parents $\mathbf{Pa}_{i}$ in the graph $\mathcal{G}$, such that
\begin{equation}  \label{BN_factorization}
    P(X) = \prod_{i=1}^{n} P(X_{i} \mid \mathbf{Pa}_{i}).
\end{equation}

Learning a Bayesian network requires estimating both components: parameters $\Theta$ and structure $\mathcal{G}$. Maximizing or marginalizing the parameters for a given structure will provide a score for each DAG. In \pkg{BiDAG}, the score $S$ of each DAG is proportional to its posterior probability given the data $D$. For computational feasibility of the implementation, it is essential that the score function factorizes into a product where each term depends only on one node and its parents:
\begin{equation}
P(\mathcal{G} \mid D)\propto P(D\mid \mathcal{G})P(\mathcal{G})=\prod_{i=1}^{n}S(X_{i},\mathbf{Pa}_{i}\mid D).
\label{scoredecomp}
\end{equation}

See \cite{BG1995} for the technical conditions guaranteeing the desired score factorization. Two score functions $S$ meeting the conditions for the decomposition in (\ref{scoredecomp}) are implemented in \pkg{BiDAG}: (1) the Bayesian Dirichlet equivalent score (BDe) \citep{BG1995} with a Dirichlet parameter prior for binary and categorical data,  and (2) the Bayesian Gaussian equivalent score (BGe)  \citep{GH2002,BGcorr1,BGcorr} with an inverse Wishart prior for continuous data.  \par

Learning the structure component of a network $\mathcal{B}$ requires finding the DAG $\mathcal{G}$, which best fits a data set $D$. As far as search-and-score methods are concerned, this means finding a graph with a score larger or equal than any other. In situations where several structures achieve similar scores, focusing on a single structure may be misleading \citep{bayes2003}. The MCMC methods in \pkg{BiDAG} account for structure uncertainty by sampling DAGs from the posterior distribution given the data $D$. Rather than examining the highest scoring DAG, we can select the model that consists only of edges whose posterior probabilities are higher than a desired threshold. Although not guaranteed to give a DAG, simulation studies show a reduction in the number of false-positive edges with this approach compared to choosing one maximally scoring model \citep{plus1}, while hardly ever resulting in directed cycles.\par

\subsection{Order MCMC} 
\label{sec:mcmc}
The rationale behind MCMC schemes is to construct a Markov chain $\mathcal{M}$ such that its stationary distribution equals the posterior distribution $P(\mathcal{G}\mid D)$ we would like to sample from. One of the schemes implemented in \pkg{BiDAG} is order MCMC, which does not operate directly on the space of DAGs but on the smaller space of orders. The posterior landscape is smoother in the space of orders than in the space of DAGs. Consequently order MCMC can achieve faster convergence with respect to structure MCMC \citep{bayes2003}.\par

A permutation $(i_1, i_2, \dots, i_n)$ of the $n$ nodes of a DAG defines a linear order $i_1 \prec i_2 \prec \dots \prec i_n$. A DAG $\mathcal{G}$ is compatible with an order $\prec$ if $i \prec j$ whenever $j$ is a parent of $i$ in $\mathcal{G}$ \citep{plus1}. We denote with $\mathbf{\Gamma_{\prec}}$ the set of all DAGs compatible with $\prec$. Each order is assigned a score that equals the sum of the scores of all DAGs compatible with this order,
\begin{equation}
R(\prec\mid D)=\sum_{\mathcal{G}\in\mathbf{\Gamma_{\prec}}}P(\mathcal{G}\mid D)\propto\sum_{\mathcal{G}\in\mathbf{\Gamma_{\prec}}}\prod_{i=1}^{n}S(X_{i},\mathbf{Pa}_{i}\mid D).
\label{orderscorefirst}
\end{equation} 
As discussed in \cite{bayes2003} we can exchange the product and sum and get the order score by summing over all parent sets compatible with the order instead of summing over all possible DAGs. Similarly to DAGs we can formulate the compatibility requirement for parent sets. A parent set $\textbf{Pa}_{i}$ of a node $i$ is compatible with an order $\prec$ if $i \prec j$ for all parents $j \in \textbf{Pa}_{i}$. For each node $i$, we denote the set of all parent sets compatible with $\prec$ by $\mathbf{U}_{\prec,i}$. Then
\begin{equation}
R(\prec\mid D)\propto\prod_{i=1}^{n}\sum_{\mathbf{Pa}_{i}\in\mathbf{U}_{\prec,i}}S(X_{i},\mathbf{Pa}_{i}\mid D).
\label{orderscore}
\end{equation} 

To construct a Markov chain in the space of orders we use the following moves from an order $\prec$ to a new order $\prec'$:
\begin{itemize}
\item Local move: swapping adjacent nodes in $\prec$
\item Global move: swapping two random nodes in $\prec$ 
\item Node relocation: in this move we place a single node $i_{k}$ in each possible position $(1,2, ...,n)$ of the current order $(i_{1}, i_{2},..., i_{n})$, while keeping the order of the other nodes in $\prec$ fixed. All $n$ orders corresponding to all possible positions of the node $i_{k}$ in the order are scored according to Equation \ref{orderscore} and the new order $\prec'$ is sampled according to these scores.
\label{ordermoves}
\end{itemize}\par
The Metropolis-Hastings acceptance probability for the first two moves is
\begin{equation}
\rho=\min\left\{1,\,\frac{R(\prec^{'} \mid D)}{R(\prec \mid D)}\right\}.
\label{acceptprob}
\end{equation}
The last move is always accepted, but it can return the current order.  \par

Order MCMC produces a sample of orders. Obtaining a sample of graphs from the posterior distribution requires an additional step of sampling DAGs from these orders according to their scores. Due to score decomposability we can do this on a per-node basis: We sample a parent set for each node from the set of parents compatible with the order, independently of other nodes. In this way, we obtain a sample of DAGs $\mathcal{G}_{1},...,\mathcal{G}_{M}$, from which we usually exclude the first $m$ to account for the burn-in period. Assuming that the Markov chain has converged within $m$ steps, we can approximate the posterior probability of any structural feature $f$ by the sample average
\begin{equation}
P(f \mid D) \approx 
\frac{1}{M-m}\sum_{i=m+1}^{M}f(\mathcal{G}_{i}),
\label{postedges}
\end{equation}
\par
where $f(\mathcal{G}_{i})$ equals 1 if the feature $f$ is present in structure $\mathcal{G}_{i}$ and 0 otherwise. \par

\subsection{Partition MCMC} 
\label{sec:partmcmc}

One advantage of the order MCMC sampling scheme compared to structure MCMC resides in its increased efficiency. Another characteristic of order MCMC is that it imposes a non-uniform prior over structures by over-representing DAGs that belong to several orders \citep{bayes2003}. To achieve unbiased sampling, \citet{Partition} proposed an MCMC scheme in the space of ordered partitions instead.  \par
A labelled partition $\Lambda$ is defined by two components: a node ordering $\prec$ and a vector of sizes of the parts $\kappa=(k_{1},...,k_{p})$, where $1\leq p\leq n$ and $\sum_{i=1}^{p}k_{i}=n$. The vector $\kappa$ divides the order $\prec$ into $p$ parts: $v_{1},...,v_{p}$, such that $v_{1}$ includes the first $k_{1}$ nodes of the permutation $\prec$, $v_{2}$ includes the following $k_{2}$ nodes, etc. A DAG $\mathcal{G}$ is compatible with a partition $\Lambda=(\prec,\kappa)$ if the following conditions are satisfied for every node $X_i$ with, say, $X_{i} \in v_{j}$:
\begin{itemize}
\item if $j<p$, $X_{i}$ has at least one parent in the part $v_{j+1}$
\item all nodes in $\textbf{Pa}_{i}$ belong to the parts with indices higher than $j$ 
\item $\textbf{Pa}_{i}=\emptyset$ if and only if $j=p$
\end{itemize}

\pkg{BiDAG} implements the following moves in the space of partitions:
\begin{itemize}
 \item swap any two nodes from different parts
 \item swap any two nodes in adjacent parts
 \item split a part or join two parts
 \item move a single node into an existing part or form a new part with the single node
 \label{partitionmoves}
\end{itemize}

The unbiased sampling with partition MCMC comes at the cost of a higher complexity for computing score tables and slower convergence as compared to order MCMC. However, the bias of order MCMC may not be a strong limitation in practice. \cite{plus1} have shown in simulation studies that the models obtained via averaging over the sample of DAGs obtained by the order MCMC scheme are very close to the ground truth structures.

 \par

\subsection{MAP discovery} 
In addition to sampling from the posterior distribution, we can also use the algorithms implemented in \pkg{BiDAG} to search for a MAP graph \citep{plus1}. To do so, we replace the sum in Equation~\ref{orderscore} with a maximum. Then, the order score equals the score of a maximum scoring DAG compatible with this order,
\begin{equation}
Q(\prec \mid D)=\prod_{i=1}^{n}\max_{\textbf{Pa}_{i}\in \mathbf{U}_{i,\prec}}S(X_{i},\textbf{Pa}_{i} \mid D)=\max_{\mathcal{G}\in\mathbf{\Gamma}_{\prec}}P(\mathcal{G} \mid D).
\label{orderscoremax}
\end{equation}

\par
In \pkg{BiDAG}, we implement order MCMC for both sampling and MAP estimation and partition MCMC only for sampling.

\subsection{Hybrid sampling scheme} 
\label{sec:plus1}
Even in the efficient order score decomposition of Equation (\ref{orderscore}), the number of possible parent sets, which need to be scored, is exponential of order $O(2^{n-1})$. To apply the algorithm to networks with, say, $n>20$ nodes, we prune the search space. \pkg{BiDAG} implements the hybrid approach of \cite{plus1} which limits the search space by means of a (possibly undirected) graph $\mathcal{H}$, whose maximal parent set  size per node is $K$, so that the number of possible parent sets reduces to $O(n2^K)$. Since we wish to sample DAGs from the posterior distribution, prior knowledge together with evidence from the data drive the pruning process to ensure that the search space $\mathcal{H}$ captures the bulk of the posterior weight. In \pkg{BiDAG}, we used the constraint-based PC algorithm \citep{PC2000} to define the search space.\par

The PC algorithm starts with a complete undirected graph and deletes edges based on conditional independence tests. After deleting as many edges as possible, we identify the skeleton graph, i.e., a graphical structure where all edges are bi-directional.  Inference with the PC algorithm includes steps to direct some edges which yield a partially directed acyclic graph (CPDAG), which represents a class of equivalent DAGs. By default, we use a PC-defined skeleton as search space $\mathcal{H}$ and not the CPDAG to avoid mistakes in directing edges.

An essential feature of \pkg{BiDAG} is the possibility to improve the initially defined search space $\mathcal{H}$ \citep{plus1}. Errors in the statistical tests of the PC algorithm can lead to the deletion of true positive edges or edges appearing in high-scoring DAGs. Simulation studies show that the true positive rate (TPR) of structures estimated by the PC algorithm decreases when the density of the ground truth DAG, defined as an average number of parents of one node, increases \citep{Kalisch2007}. If the search space $\mathcal{H}$ lacks some of the edges from a MAP DAG $\mathcal{G}^{MAP}$, we will not be able to find it when searching in $\mathcal{H}$.  To address this limitation, \cite{plus1} propose to expand the search to an extended space $\mathcal{H}^{+}$ in which the possible parent sets of every node include not only all combinations of parents of $X_{i}$ in $\mathcal{H}$ but also these parent sets joined with any other node that is not a parent of $X_{i}$ in $\mathcal{H}$. Searching in $\mathcal{H}^{+}$ provides the opportunity to correct for any mistakes of the pruning algorithm and yields higher scoring DAGs. We will refer to $\mathcal{H}$ as the core search space and to $\mathcal{H}^{+}$ as the extended search space. \par

The improvements we can achieve by simply searching in the extended space are limited. For example, if two or more parents are missing in the same node's parent set, the approach would allow us to recover only one of them. However, if we iterate the procedure we may be able to correct for more than one mistake per parent set. The iterative order MCMC procedure is summarized in Algorithm \ref{algo:it} below: \par

\begin{algorithm}
\KwIn{data $D$}
\KwOut{MAP estimate $\mathcal{G}^{ \mathrm{max}}$, \ optimized search space $\mathcal{H}$}
Initiate the search space $\mathcal{H}$ with the PC algorithm or an arbitrary adjacency matrix\\
Run the order MCMC scheme on the search space $\mathcal{H}^{+}$\\
Assign $\mathcal{G}^{\mathrm{max}}$ the maximally scoring DAG obtained by the MCMC scheme \\
Update $\mathcal{H}$, $\mathcal{H}^{\mathrm{old}}$: \linebreak 
$\mathcal{H}^{\mathrm{old}}=\mathcal{H}$
\linebreak
$\mathcal{H}=\mathcal{H}\cup\mathcal{G}^{\mathrm{max}}$ \\
Repeat Steps 2 to 4, till $\mathcal{H}= \mathcal{H}^{\mathrm{old}}$\\

\caption{iterative MCMC procedure}
\label{algo:it}
\end{algorithm}

Simulation studies show that the iterative MCMC procedure can improve even poor search spaces containing only $50-60\%$ true positive edges so that the final space contains $90-100\%$ true positive edges \citep{plus1}. However, the worse the original search space is, the more expansion iterations we need to optimize it. Defining a reasonable search space $\mathcal{H}$ to start with can significantly decrease the total runtime of the iterative MCMC scheme. In the current version of \pkg{BiDAG}, in addition to the PC algorithm, it is possible to define the search space with an arbitrary adjacency matrix, which may stem from expert knowledge or another algorithm for structure learning. Computational complexity of the MCMC schemes in $\mathcal{H}^{+}$ is higher than in  $\mathcal{H}$. We will discuss differences in runtimes between using $\mathcal{H}$ and $\mathcal{H}^{+}$ in more detail in Section \ref{runtime}.

%% file: chapters/Package.tex
\section{BiDAG package}
\label{bidag}

The core functions of the package \imcc, \omcc, \code{partitionMCMC} can be used for structure learning and sampling of Bayesian networks and dynamic Bayesian networks. The remaining functions can be dividend into four main groups: convergence diagnostics, model averaging, model comparison, and network visualization. In this section, we describe the most important functions from all groups.

\subsection{Constructing the score object}

All functions for structure learning require an object of class \code{score\-parameters}, which stores the data and other quantities needed to score Bayesian networks. We can construct an object of class \code{score\-parameters} using the function
\begin{CodeChunk}
\begin{CodeInput}
scoreparameters(scoretype = c("bge", "bde", "bdecat", "usr"), data, 
bgepar = list(am = 1, aw = NULL),
bdepar = list(chi = 0.5, edgepf = 2),
bdecatpar = list(chi = 0.5, edgepf = 2),
dbnpar = list(samestruct = TRUE, slices = 2, b = 0), 
usrpar = list(pctesttype = c("bge", "bde", "bdecat")), 
DBN = FALSE, weightvector = NULL, bgnodes = NULL, edgepmat = NULL, 
nodeslabels = NULL)
\end{CodeInput}
\end{CodeChunk}
The \code{data} should be in the form of a \code{data.frame} or a \code{matrix} with $N$ rows and $n$ columns, where $n$ is the number of variables in the Bayesian network and $N$ the number of observations. The parameter \code{scoretype} defines which score function is used: \code{bde} for binary data, \code{bdecat} for categorical data, \code{bge} for continuous data and \code{usr} for a user-defined score. An optional parameter \code{weightvector} defines the weight of each observation. The need for weighted data may arise, for example, in survey analysis \citep{Kuipers2018s} and Bayesian network-based clustering \citep{Kuipers2018}.\par

There are several ways to include prior information in structure learning. The parameter \code{bgnodes} lists root nodes (those who can have children but no parents). For example, we may expect that the gender of a participant in a survey data can have an effect on the answers, but not the opposite. Through the parameter \code{edgepmat} we can choose to penalize particular edges in the search space: we do not exclude them completely but simply reduce their chance to be sampled. Excluding the edges from the search space is also possible via the parameter \code{blacklist} of the structure learning functions, which we will discuss in the next section. 

\subsection{Structure learning and sampling}
\label{sec:package}
The functions \omc and \code{partitionMCMC} implement the order and partition MCMC schemes, respectively. While \omc can perform both sampling from the posterior and MAP discovery, \code{partitionMCMC} is limited to sampling from the posterior. The latter scheme is the only efficient option in \pkg{BiDAG} to get a sample from the posterior using a uniform prior over structures. The function \imc implements the iterative approach \citep{plus1} described in Section \ref{sec:plus1} for optimizing the search space or MAP DAG discovery. \par

All three functions for structure learning share a similar syntax. We explain in detail the important parameters of the function
\begin{CodeChunk}
\begin{CodeInput}
orderMCMC(scorepar, MAP = TRUE, plus1 = TRUE, chainout = FALSE,
scoreout = FALSE, moveprobs = NULL, iterations = NULL, stepsave = NULL,
alpha = 0.05, cpdag = FALSE, gamma = 1, hardlimit = ifelse(plus1, 14, 20),
verbose = FALSE, startspace = NULL, blacklist = NULL,
startorder = NULL, scoretable = NULL)
\end{CodeInput}
\end{CodeChunk}
and then describe the differences for \code{partitionMCMC} and \imcc. All of them  only require the parameter \code{scorepar}, a \code{scoreparameters} object as described in the previous section. All other parameters are either optional or have default values. However, the MCMC schemes are very flexible, and the parameters should be consistent with the objectives and constraints of a particular structure learning problem. Parameters fall into four categories: 
\begin{itemize}
\item parameters defining the search space: \code{startspace}, \code{cpdag}, \code{plus1}
\item parameters of the Markov chain: \code{MAP}, \code{iterations}, \code{stepsave}, \code{moveprobs}
\item parameters to include prior information: \code{blacklist}, \code{startorder}
\item parameters defining objects included in the output: \code{chainout}, \code{scoreout}
\end{itemize}

The \code{MAP} parameter defines whether we wish to use order MCMC for finding a MAP DAG or for sampling from the posterior. For the latter, we should set \code{MAP} \code{FALSE}, in which case the order score is calculated according to Equation (\ref{orderscore}), otherwise using Equation (\ref{orderscoremax}). At each MCMC step, if \code{MAP} equals \code{TRUE}, the algorithm returns the maximally scoring DAG from the order. Otherwise, it samples DAGs from the orders according to their scores. \par

The number of MCMC iterations defined by the parameter \code{iterations} should be large enough for the MCMC chain to converge, while still controlling the runtime. The number of MCMC iterations required for convergence cannot be calculated analytically. Heuristics \citep{Partition} and simulation studies \citep{plus1} suggest that we need $O(n^{2}\log{n})$ iterations to reach convergence or discover a maximum DAG. Motivated by this finding, in \pkg{BiDAG}, we set the default value of \code{iterations} to $6n^{2}\log{n}$ for order MCMC and $20n^{2}\log{n}$ for partition MCMC. \par

To avoid excessively long runtimes, the algorithm does not sample DAGs at each MCMC iteration but once every \code{stepsave} steps. The idea of \code{stepsave} is that the number of iterations needed for the MCMC chain to converge is large and can be tens of thousands or even millions, while the required number of DAGs sampled from the posterior is usually much smaller. Sampling DAGs from the orders after each $l$ steps significantly reduces the runtime without having a negative effect on convergence.  By default, we define \code{stepsave} in such a way that the algorithm samples $1001$ DAGs.\par

The parameter \code{startspace} can define the search space via a binary adjacency matrix of size $n\times n$. An entry $[i,j]$ in the adjacency matrix is $1$ to indicate the presence of an edge from $i$ to $j$, and it is zero otherwise. The search space can be an arbitrary graph without the acyclicity requirement. When edges are  bidirectional both entries $[i,j]$ and $[j,i]$ should be equal to 1. Unit entries in column $j$ determine the permissible parent sets for node $j$. When \code{startspace} is not specified, we define the search space by the skeleton estimated by the PC algorithm or by an equivalence class represented by a CPDAG if the parameter \code{cpdag} equals \code{TRUE}. \par

The parameter \code{alpha} defines the significance level $\alpha$ used in the conditional independence tests of the PC algorithm. Larger $\alpha$ values lead to larger search spaces, which decrease the risk that true positive edges are absent as a result of errors in the statistical tests. By the same principle though, high $\alpha$ values will also increase the number of false-positive edges. While a higher number of false positive edges in the search space does not affect the goodness of fit of the resulting structures, it can negatively affect the runtime. Moreover, larger $\alpha$ values also imply longer runtimes for the PC algorithm, which is worst-case exponential. By default $\alpha=0.05$. \par

The parameter \code{plus1} indicates whether the algorithm should perform the search in the core search space $\mathcal{H}$ or in the extended space $\mathcal{H}^{+}$. When \code{plus1} equals \code{TRUE}, then the chain is constructed in $\mathcal{H}^{+}$ instead of $\mathcal{H}$, as described in Section \ref{sec:plus1}.\par

The parameter \code{blacklist} defines all single edges we wish to remove from the search space, and hence they will not appear in any of the sampled DAGs. If a node is not allowed to have any parents, it is computationally more efficient to define it as a background (root) node via the parameter \code{bgnodes} in the \code{scorepar} object instead of specifying all edges from any other node in a \code{blacklist}. The parameter \code{edgepmat} of the function \code{scoreparameters} mentioned above can be regarded as a soft version of the blacklist. \par

The function \imc implements the iterative order MCMC scheme and thus inherits most of the parameters from the function \omcc. It includes additional parameters to define the iterative expansion of the search space. \code{plus1it} defines the number of iterations of expansion of the search space; when \code{plus1it} is not specified, the search space expands until no edges can be added to the search space to improve the score of a maximally scoring DAG. Other parameters in \imc define limits on the maximum number of edges we can add to the search space. When extending the search space, the maximal parent set size $K$ may increase as well. In Section \ref{runtime}, we discuss how  $K$  affects the runtime. The parameter \code{hardlimit} sets a limit on the number of parents any node may have. When we hit the limit for one node, the algorithm prevents adding further elements to that node's parent set, but it can still expand the parent sets of other nodes until they all reach the limit or the score does not improve further.  Another parameter controlling the expansion of the search space is \code{mergetype}. The possible values of \code{mergetype}, namely \code{dag}, \code{cpdag}, and \code{skeleton}, correspond to merging the core space $\mathcal{H}$ with a maximally scoring graph $\mathcal{G}^{\max}$, its equivalence class or a skeleton accordingly. \par

\imc also accepts the option \code{MAP=FALSE} to define the expansion graph $\mathcal{G}^{*}_{i}$ at each iteration on the basis of a sample of graphs and a posterior probability threshold given by the parameter \code{posterior}. $\mathcal{G}^{*}_{i}$ includes all edges with posterior probabilities higher than the threshold. \par

By default, only MAP DAGs are stored at each search space expansion step in \imcc. However, one may want to inspect DAGs other than the maximum. The element  \code{addtrace} of the \code{iterativeMCMC} object stores adjacency matrices of all sampled DAGs from all MCMC steps when the parameter \code{chainout} set to TRUE.

The function \code{partitionMCMC} has a similar structure to \omcc, but it does not use the parameters \code{MAP} and \code{plus1}, since it only samples from the posterior in the extended search space. When the parameter \code{startspace} is not defined, by default, the procedure defines the first search space via the PC algorithm and then improves it by \imcc. \par 

\subsection{Bayesian model averaging and posterior model selection}

To calculate posterior probabilities of single edges based on a sample of graphs from MCMC schemes we can use the function

\begin{CodeChunk}
\begin{CodeInput}
edgep(MCMCchain, pdag = FALSE, burnin = 0.2, endstep = 1)
\end{CodeInput}
\end{CodeChunk}

where the parameter \code{MCMCchain} is an object of class \code{orderMCMC} or \code{partitionMCMC}. The parameter \code{burnin} defines the proportion of samples to discard as burn-in. We can also perform posterior model selection by constructing a graph consisting only of edges with posterior probability higher than a certain threshold with the function 
\begin{CodeChunk}
\begin{CodeInput}
modelp(MCMCchain, p, pdag = FALSE, burnin = 0.2)
\end{CodeInput}
\end{CodeChunk}
which however is not guaranteed to result in a DAG.
When building a consensus graph from a sample of DAGs it is possible to account for the uncertainty related to equivalence class by setting the parameter \code{pdag} to \code{TRUE}. In this case, we first convert all DAGs in the sample to CPDAGs corresponding to their equivalence classes.

\subsection{Diagnostic plots}
The convergence of the MCMC schemes is essential both for sampling from the posterior distribution as well as for MAP discovery. It is generally impossible to prove that the Markov chain has converged. However, diagnostics plots may help analyzing convergence and spotting cases when convergence was not reached. Trace plots are the basic tool for convergence diagnostics.  For objects of classes \code{orderMCMC}, \code{partitionMCMC} and \code{iterativeMCMC}, the method \code{plot} is available, which plots the trace of log scores of sampled DAGs. \par

To plot the changes in posterior probabilities of all single edges with the addition of new graphs from the sample according to Equation (\ref{postedges}) we can use the function
\begin{CodeChunk}
\begin{CodeInput}
plotpedges(MCMCtrace, cutoff = 0.2, pdag = FALSE, onlyedges = NULL, 
highlight = NULL, ...)
\end{CodeInput}
\end{CodeChunk}

Large fluctuations of posterior probabilities are possible at the beginning, but while approaching convergence posterior probabilities should also reach stable levels. \par

Convergence diagnostic plots based on a single chain may be misleading. For a better understanding of convergence we can examine jointly several independent MCMC runs with random starting points. If all chains converge, the DAGs in each chain should represent the posterior distribution in a similar way. Posterior probabilities of single edges calculated on the basis of each sample should then be close to each other. If some chains do not converge, we are likely to see significant differences between posterior probabilities of single edges. We can plot the concordance between pairs of MCMC runs using the function

\begin{CodeChunk}
\begin{CodeInput}
plotpcor(pmat, highlight = 0.3, printedges = FALSE, cut = 0.05, ...)
\end{CodeInput}
\end{CodeChunk}

where the parameter \code{pmat} is a list of matrices containing posterior probabilities of single edges; such a list can be created by applying the function \code{edgep} to a list of objects of class \code{orderMCMC} or \code{partitionMCMC}. We can also inspect the edges whose posterior probabilities differ by more than \code{highlight} in the first two matrices by setting \code{printedges} to TRUE.

\subsection{Model comparison}
The function \code{DAGscore}  computes the score of a single DAG. When the goal is MAP discovery, we can use this function to compare structures estimated by different algorithms. We can also compare scores of the estimated structures to the score of the ground truth DAG when the latter is known. \par

To compare the performance of structure learning algorithms it is useful to assess how close the estimated structure is to the ground truth DAG on the basis of a certain distance measure. The function \code{compareDAGs} allows several measures: the number of false-positive edges (FP), the number of false-negative edges (FN), the true positive rate (TPR), the structural Hamming distance (SHD) and others. All measures apart from SHD refer to differences in the skeletons of two DAGs, i.e., the directions of the edges are disregarded. SHD equals the sum of all types of mistakes: false negatives, false positives, and edges with erroneous directions. The functions \code{plotdiffs}, \code{plotdiffs.DBN} and \code{plot2in1} can be used to visualize the differences and similarities between two graphs.\par

%% file: chapters/DBNs.tex
\section{Structure learning of dynamic Bayesian networks}

\label{DBNs}
\begin{figure}[h!]
\centering
\includegraphics{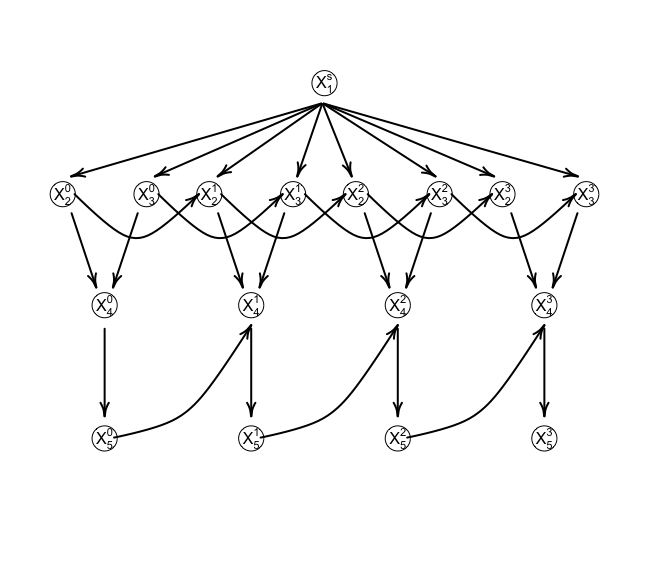}
\caption{Unfolded structure of a first-order DBN consisting of four time slices. Each time slice includes one static variable $X_{1}^{S}$ and four dynamic variables $X_{2}^t$, $X_{3}^t$, $X_{4}^t$, $X_{5}^t$, for $t$ = 0, 1, 2, 3.}
\label{fullDBN}
\end{figure}

\begin{figure}[t!]
     \centering
     \begin{subfigure}[b]{0.18\textwidth}
         \centering
         \includegraphics[width=\textwidth]{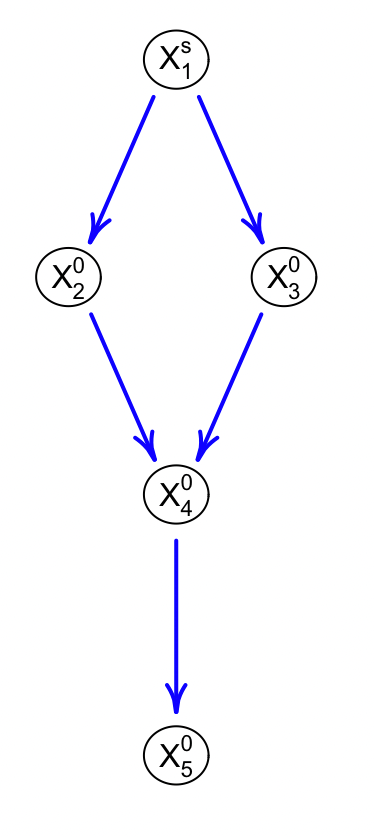}
         \caption{$\mathcal{G}_{0}$}
         \label{fig:DBNstrinit}
     \end{subfigure}
     \hfill
     \begin{subfigure}[b]{0.54\textwidth}
         \centering
         \includegraphics[width=\textwidth]{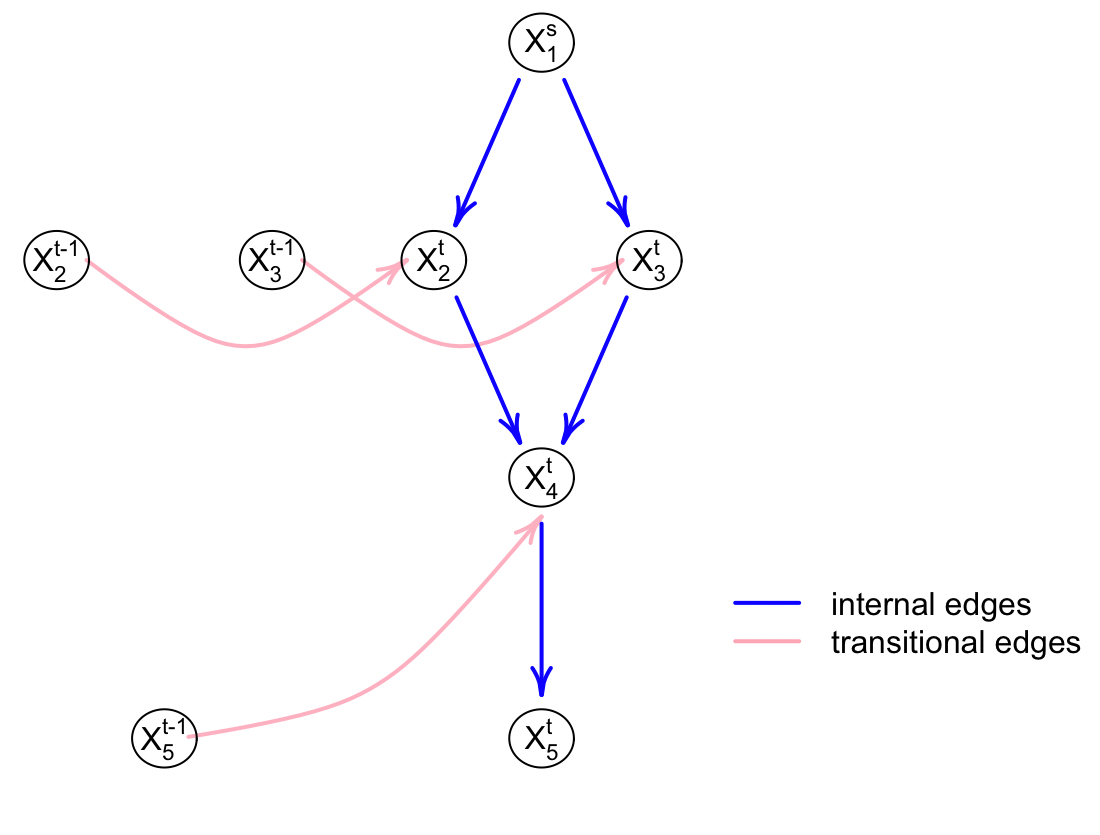}
         \caption{$\mathcal{G}_{\rightarrow}$}
         \label{fig:DBNstrtrans}
     \end{subfigure}
   \caption{Initial $\mathcal{G}_{0}$ and transition $\mathcal{G}_{\rightarrow}$ structures representing the first-order DBN whose unfolded structure is depicted in Figure \ref{fullDBN}.}
        \label{compactDBN}
\end{figure}

A dynamic Bayesian network (DBN) is a graphical model that encodes temporal relationships between random variables in $\textbf{X}$. A DBN defines a joint probability distribution over $\textbf{X}^t = (X^t_{1}, \dots, X^t_{n})$ for all discrete time points $t = 1, \dots, T$. The random variable $X^t_i$ describes feature $i$ at time point $t$. In \pkg{BiDAG}, we consider first-order homogeneous DBNs, where the conditional probability distributions $P(\textbf{X}^{t} \mid \textbf{X}^{t-1})$ are assumed to be the same for all time points $t$. In a first-order DBNs, variables in time slice $t$ can only depend on other variables in the same time slice or on variables in the previous time slice $t-1$. The structure of a first-order homogeneous DBN $\mathcal{G}$ is fully specified by the initial structure $\mathcal{G}_{0}$ and the transition structure $\mathcal{G}_{\rightarrow}$. $\mathcal{G}_{0}$ represents the structure of the first slice, and $\mathcal{G}_{\rightarrow}$ represents the structure for transitioning between any pair of consecutive time slices. Specifically, $\mathcal{G}_{\rightarrow}$ includes internal edges, i.e., edges between the nodes within the same time slice and transition edges,  i.e., edges from the nodes in a previous time slice to the current time slice. The unfolded DBN structure $\mathcal{G}$ shown in Figure \ref{fullDBN} can be more compactly represented as the two structures $\mathcal{G}_{0}$ and $\mathcal{G}_{\rightarrow}$ shown in Figure \ref{compactDBN}.

\pkg{BiDAG} can also be used for learning DBNs from data.
When we initialize the score object with the function \code{scoreparameters} we set the parameter \code{DBN} to \code{TRUE} (the default is \code{FALSE}). The \code{data} object must adhere to a special DBN format to perform structure learning correctly.  The number of columns must equal the number of variables in all time slices $b+n\cdot T$, where $b$ is the number of static variables, $n$ the number of dynamic variables in one time slice and $T$ the number of time points. All $b$ static variables, if present, have to be in the first $b$ columns of the \code{data} and $b$ should be specified via the parameter \code{dbnpar}, containing a list of variables specific to DBNs. We assume that static variables are present in every time slice, but since they do not change over time we need to store their values only once. The next $n \cdot T$ columns should store the observations of the dynamic variables over all time slices. They need to be ordered in such a way that for each group of $n$ variables, the $i^{\rm th}$ column of group $t$ contains the observations of the variable $X_{i}^{t}$. \par

In \pkg{BiDAG}, we consider a special case when the structure within the first time slice is the same as the internal structure in all other time slices, which we indicate by setting to \code{TRUE} the slot \code{samestruct} in the parameter \code{dbnpar}. Otherwise we learn the initial and transition structures independently. \par

%% file: chapters/Examples.tex
\section{Examples on simulated data}
\label{examples}
We consider two data sets to demonstrate possible ways of working with the \pkg{BiDAG} package. The first simulated dataset is \code{gsim100}; it includes 100 observations generated from a randomly generated DAG with $n=100$ nodes, corresponding to Gaussian random variables. The second simulated dataset, \code{DBNdata}, contains observations from five consecutive time points of a DBN consisting of 12 dynamic 3 static variables.

\subsection{MAP discovery} 
\label{sec:map}
We first demonstrate how to use the algorithms in \pkg{BiDAG} for MAP discovery, which we can perform via the functions \omc and \imcc. Both functions rely on the order MCMC scheme, but they use different approaches to construct the search space. \par

To run any of the implemented MCMC schemes, we need to construct an object of class \code{scoreparameters}.
\begin{CodeChunk}
\begin{CodeInput}
R> score100 <- scoreparameters("bge", gsim100)
\end{CodeInput}
\end{CodeChunk}

We first learn the MAP network from \code{gsim100} dataset  by running \code{orderMCMC} on a search space defined by the PC algorithm. It is the least computationally expensive of all options to define a search space but also prone to mistakes. 

\begin{CodeChunk}
\begin{CodeInput}
R> basefit <- orderMCMC(scorepar = score100, MAP = TRUE, plus1 = FALSE)
\end{CodeInput}
\end{CodeChunk}
The score of the maximum DAG sound in the core search space is lower than the score of the ground truth structure:

\begin{CodeChunk}
\begin{CodeInput}
R> basefit$score
R> DAGscore(scorepar = score100, incidence = gsimmat)
\end{CodeInput}
\begin{CodeOutput}
[1] -17947.39
[1] -15239.79
\end{CodeOutput}
\end{CodeChunk}
\par

By looking at structural differences, we can see that most differences in the estimated DAG come from the low number of discovered true-positive edges:

\begin{CodeChunk}
\begin{CodeInput}
R> compareDAGs(basefit$DAG, gsimmat)[c("TPR", "FPRn", "SHD")]
\end{CodeInput}
\begin{CodeOutput}
 TPR  FPRn   SHD 
 0.58  0.03 97.00
\end{CodeOutput}
\end{CodeChunk}

The TPR of the highest scoring graph found in the core search space is only 58\%. In an attempt to improve the search space and estimate a better DAG, we use the \imc procedure. We do not set any limit with the parameter \code{plus1it} and let the algorithm expand the search space until no additional edges can improve the score of the maximum DAG found.

\begin{CodeChunk}
\begin{CodeInput}
R> iterativefit <- iterativeMCMC(score100,  scoreout = TRUE, verbose = FALSE)
\end{CodeInput}
\end{CodeChunk}

For each expansion iteration, the algorithm constructs a new MCMC chain, and the scheme may take a while to run. When the parameter \code{verbose} equals TRUE, messages in the output indicates the iteration currently running. \par

\begin{CodeChunk}
\begin{CodeInput}
R> summary(iterativefit)
\end{CodeInput}
\begin{CodeOutput}
object of class 'iterativeMCMC'

Results: 
maximum score DAG with 100 nodes and  198  edges: 
maximum DAG score= -15195.82 

algorithm: iterative order MCMC 
number of search space expansion steps: 7 
number of edges in the initial search space: 204 
number of added edges: 194 
total number of MCMC iterations: 1127000 
total number of MCMC sampling steps (length of trace): 7007 
number of MCMC iterations per expansion step: 161000 
number of MCMC sampling steps per expansion step: 1001 
initial search space: PC 
sample/MAP:  MAP 

Additional output: 
scoretable, object of class 'scorespace'   
\end{CodeOutput}
\end{CodeChunk}

\begin{figure}[b!]
\centering
\includegraphics{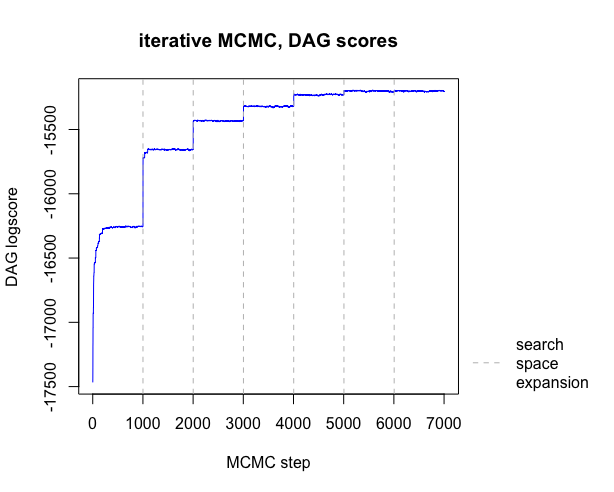}
\caption{Trace plot of saved DAG scores obtained by the function \code{iterativeMCMC} with the parameter \code{MAP=TRUE}.}
\label{fig:traceit}
\end{figure}

The iterative order MCMC scheme added 194 edges to the initial PC-defined search space in 7 iterations. We can observe how the score improved with each search space expansion step by looking at the trace plot depicted in Figure  \ref{fig:traceit}:

\begin{CodeChunk}
\begin{CodeInput}
R> plot(iterativefit)
\end{CodeInput}
\end{CodeChunk}

The scores of the DAGs sampled at the final expansion step improved significantly compared to the initial step. Moreover, the score of a MAP DAG found in the last iteration of \code{iterativeMCMC} (-15195.82) is higher than the score of the ground truth structure (-15239.79), and much higher than the score of a MAP graph found in the core PC-defined search space (-17947.39). Outside of simulation studies, DAG score is the most used criterion that informs model selection and \code{iterativeMCMC} has shown great performance in maximizing the score \citep{plus1}. 

Since we know the ground truth DAG, we can use the function \code{itercomp} to assess how close the estimated MAP structures are to the true DAG with each expansion of the search space:

\begin{CodeChunk}
\begin{CodeInput}
R> it100 <- itercomp(iterativefit, gsimmat)
R> plot(it100, vars = c("score", "TPR"), showit = c(1:6))
R> plot(it100, vars = c("FP", "SHD"), col = 2, showit = c(1:6))
\end{CodeInput}
\end{CodeChunk}

\begin{figure}[t!]
     \centering
     \begin{subfigure}[b]{0.45\textwidth}
         \centering
         \includegraphics[width=\textwidth]{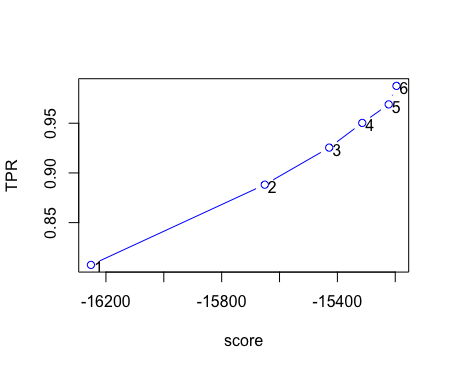}
         \caption{TPR and score}
         \label{fig:sim1}
     \end{subfigure}
     \hfill
     \begin{subfigure}[b]{0.45\textwidth}
         \centering
         \includegraphics[width=\textwidth]{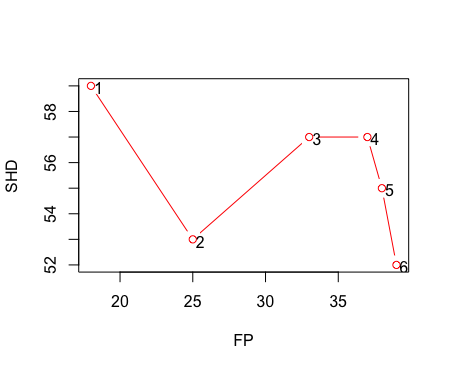}
         \caption{FP and SHD}
         \label{fig:sim2}
     \end{subfigure}
   \caption{Structure fit changes through iterative expansions of the search space: \code{iterativeMCMC} was applied to the simulated dataset \code{gsim100}. At each search space expansion, the MAP DAG is stored together with its score and compared to the ground truth structure with the function \code{itercomp}. For (a), TPR (true positive rate) and DAG score were used, while for (b), FP (number of false-positive edges) and SHD (structural Hamming distance) were used to visualize the changes.}
\label{fig:simres}
\end{figure}

As visualized in Figure \ref{fig:sim1}, the results of this comparison show that the TPR grows as the score increases with each search space expansion and is very close to 1 in the last iteration. However, the number of false-positive edges grows as well, and thus, the improvement of SHD is not as impressive as of TPR (\ref{fig:sim2}).

Simulation studies help us by guiding what we can expect from applying a particular method in a specific simulation setting, e.g., low sample size. In our example,  using the PC-defined search space results in a low TPR. While \code{iterativeMCMC} helps with the TPR and the DAG score, it does not necessarily result in the best structure fit, as previously mentioned, due to a possible increase in false positives. In the next section, we will describe how to use \pkg{BiDAG} to obtain consensus graphs that help mitigate this problem. \par

\subsection{Sampling graphs from posterior distribution} 
\label{sec:sample}
So far, we focused on finding one maximally scoring DAG. For sampling from the posterior distribution, we can use the same functions as for MAP learning by setting the parameter \code{MAP} to  \code{FALSE}. In addition, we can use the function \code{partitionMCMC} for sampling with a uniform prior over structures. For sampling, it is important that the search space includes as many true positives as possible. The iterative MCMC scheme successfully optimized the search space in multiple simulation settings \citep{plus1}. Thus, we pass to sampling function the search space previously optimized with the function \imc in Section \ref{sec:map} via the parameter \code{startspace}. 
 \begin{CodeChunk}
\begin{CodeInput}
R> orderfit <- orderMCMC(score100, MAP = FALSE, chainout = TRUE,
+ startspace = iterativefit$endspace)
R> plot(orderfit)
\end{CodeInput}
\end{CodeChunk}
 
\begin{figure}[t!]
\centering
\includegraphics{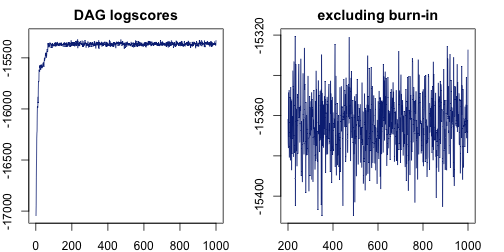}
\caption{Trace plot of DAGs sampled by the function \code{orderMCMC} with the parameter \code{MAP=FALSE}. The sampling is performed on a search space which was previously optimized by the function \code{iterativeMCMC}.}
\label{fig:exsamp}
\end{figure}

 For MCMC sampling schemes, it is important to check if the chain has converged, and we can look at diagnostic plots, which may highlight lack of convergence. The trace plot in Figure \ref{fig:exsamp} shows the scores of all sampled DAGs. When a random order is used as a starting point, typically, the scores increase sharply in the beginning, reflecting the burn-in period of the chain. A sharp increase is visible on the left subgraph, while the right subgraph shows the trace plots of scores after excluding the burn-in period. If we choose the burn-in period adequately, the scores on the right will stay in a narrow stable range. To modify the default burn-in period of 20\% we can set the parameter \code{burnin} to another value. 
 
 To demonstrate a more rigorous convergence diagnostic, we need to run MCMC two (or more) times for the same data but with different starting points. By default, the starting point of each run is random unless the parameter \code{startorder} is set to a specific value.   
 
\begin{figure}[t!]
     \centering
     \begin{subfigure}[b]{0.45\textwidth}
         \centering
         \includegraphics[width=\textwidth]{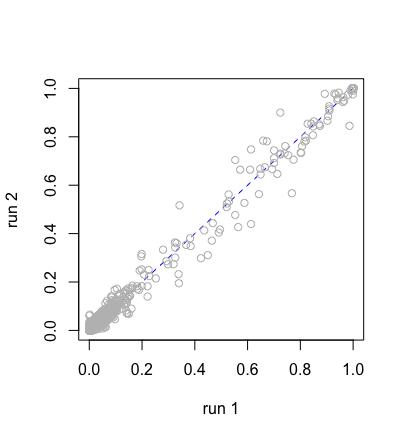}
         \caption{order MCMC (default \code{iterations})}
         \label{fig:postcomp1}
     \end{subfigure}
     \hfill
     \begin{subfigure}[b]{0.45\textwidth}
         \centering
         \includegraphics[width=\textwidth]{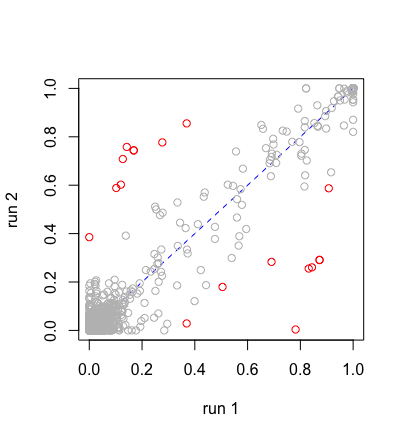}
         \caption{partition MCMC (default \code{iterations})}
         \label{fig:postcomp2}
   \end{subfigure}
   \caption{Convergence diagnostic plot: concordance of posterior probabilities estimates of single edges between pairs of MCMC runs. For each subfigure two pairs of samples of DAGs were obtained by order/partition MCMC schemes. Posterior probabilities of all edges were calculated based on each sample with the function \code{edgep} and visualized pair-wise with \code{plotpcor}. Red points correspond to edges whose posterior probabilities differ by more than 0.3 between two samples.}
\label{fig:postcomp}
\end{figure}

 We proceed with computing posterior probabilities of edges based on each of the two chains with the function \code{edgep} and visualize the results with the function \code{plotpcor}, Figure \ref{fig:postcomp2}.

\begin{CodeChunk}
\begin{CodeInput}
 R> orderfit2 <- orderMCMC(score100, MAP = FALSE, chainout = TRUE,
 + scoretable = iterativefit$scoretable)
 R> epd <- lapply(list(orderfit, orderfit2), edgep, pdag = TRUE)
 R> plotpcor(epd)
\end{CodeInput}
\end{CodeChunk}
 
 The concordance plot in Figure \ref{fig:postcomp1} does not indicate any convergence issues. All points are close to diagonal, meaning that posterior probabilities of single edges based on two samples of DAGs produced by \code{orderMCMC} are close to each other.  
 
We repeat the same procedure for a pair of runs of \code{partitionMCMC}. The concordance plot in Figure \ref{fig:postcomp2} indicates convergence issues. Several edges have a high posterior probability in one run and low in the other. When the concordance plot looks concerning, increasing the number of \code{iterations} and repeating sampling is recommended. However, even with the default number of iterations, partition sampling takes longer than order sampling, and any further increase in runtime might be unwanted. It is important to be aware of the convergence and complexity properties of both approaches in order to choose the right trade-off between runtime and prior over structures. As mentioned in section \ref{sec:partmcmc}, partition MCMC was designed in order to impose a uniform prior over structures. However, this feature comes at the cost of slower convergence and longer runtimes. In this example, we proceed with the sample obtained by the order MCMC scheme for model selection. Section \ref{sec:app} will demonstrate an example using real data, where the convergence diagnostics of the partition MCMC sample does not indicate any issues. \par

In Section \ref{sec:map}, we have compared an estimated MAP DAG to the ground truth structure by comparing their scores and skeletons. The estimated MAP DAG contains 39 false-positive edges, around 20\% of all edges in the discovered DAG. Simulation studies show that in cases when the number of observations is low, even high scoring structures may contain a lot of false-positive edges \citep{plus1}.  With a sample of DAGs from the posterior distribution, we can apply an alternative approach to model selection based on posterior probability estimates of single edges, by using the function \code{modelp} and setting the parameter \code{p} to a desired value. Since we also know the ground truth structure in this example, we can use the function \code{samplecomp} to demonstrate how models selected based on a range of posterior thresholds compare to the ground truth. We set the parameter \code{pdag} to \code{TRUE} to account for equivalence class uncertainty and accordingly, in each case, compare the chosen model to an equivalence class corresponding to the ground truth DAG:

\begin{CodeChunk}
\begin{CodeInput}
R> samplecomp(orderfit, gsimmat, pdag = TRUE, p = c(0.5, 0.7, 0.9, 0.95))
\end{CodeInput}

\begin{CodeOutput}
   TP FP FN  TPR  FPR FPRn  FDR SHD    p
1 159 28  2 0.99 0.01 0.17 0.15  36 0.50
2 158 17  3 0.98 0.00 0.11 0.10  25 0.70
3 146  4 15 0.91 0.00 0.02 0.03  19 0.90
4 140  0 21 0.87 0.00 0.00 0.00  21 0.95
\end{CodeOutput}
\end{CodeChunk}

Each row in the table corresponds to the result of comparing a consensus graph based on the posterior threshold in the last column to the ground truth CPDAG. For example, a graph consisting only of edges with an estimated probability higher than 0.90 contains only four false-positive edges, while maintaining a rather high TPR of more than 90\%.  To further show in which settings posterior model selection based on a threshold may provide an advantage over choosing one highest scoring model, we also applied a similar MCMC scheme to the dataset generated from the same network but with a larger number of observations (dataset \code{gsim}, $N=1000$). The results of comparing the estimated models with the ground truth structure are summarized in Table \ref{summary}.

\begin{table}[ht]
\centering
\begin{tabular}{rrrrrrr}
  \hline
  & \multicolumn{3}{c}{N=100} & \multicolumn{3}{c}{N=1000} \\
 & TP & FP & SHD & TP & FP & SHD \\ 
  \hline
  MAP & 159 & 39 & 52 & 161 & 8 & 12 \\ 
  $p=0.50$ & 159 & 28 & 36 & 161 & 7 & 10 \\ 
  $p=0.90$ & 146 & 4 & 19 & 160 & 4 & 5 \\ 
  $p=0.95$ & 140 & 0 & 21 & 158 & 4 & 9 \\ 
   \hline
\end{tabular}
 \caption{Comparison between MAP and  posterior threshold-based models for the two data sets \code{gsim100} and \code{gsim} generated from the same graph and containing 100 and 1000 observations accordingly.}
 \label{summary}
\end{table}

For both sample sizes, a posterior threshold-based model for $p=0.5$ has as many true edges as the MAP estimate while reducing the number of false-positive edges. A more stringent threshold can further reduce false positives. For example, for $N=100$ and $p=0.95$ there are no false-positive edges in the estimated model compared to  $20\%$ in the MAP DAG. Similarly, SHDs between consensus and ground truth models are much smaller than SHDs between MAP and the ground truth models. We can also observe that for $N=100$ the reduction in false-positive edges is more pronounced than for $N=1000$. Similar results were observed in larger-scale simulation studies by \cite{plus1}.\par

\subsection{Learning DBNs} 

All structure learning functions \omc, \imc, \code{partitionMCMC} can also be applied to structure learning and sampling of DBNs.

\begin{figure}[t!]
\centering
\includegraphics[width=\textwidth]{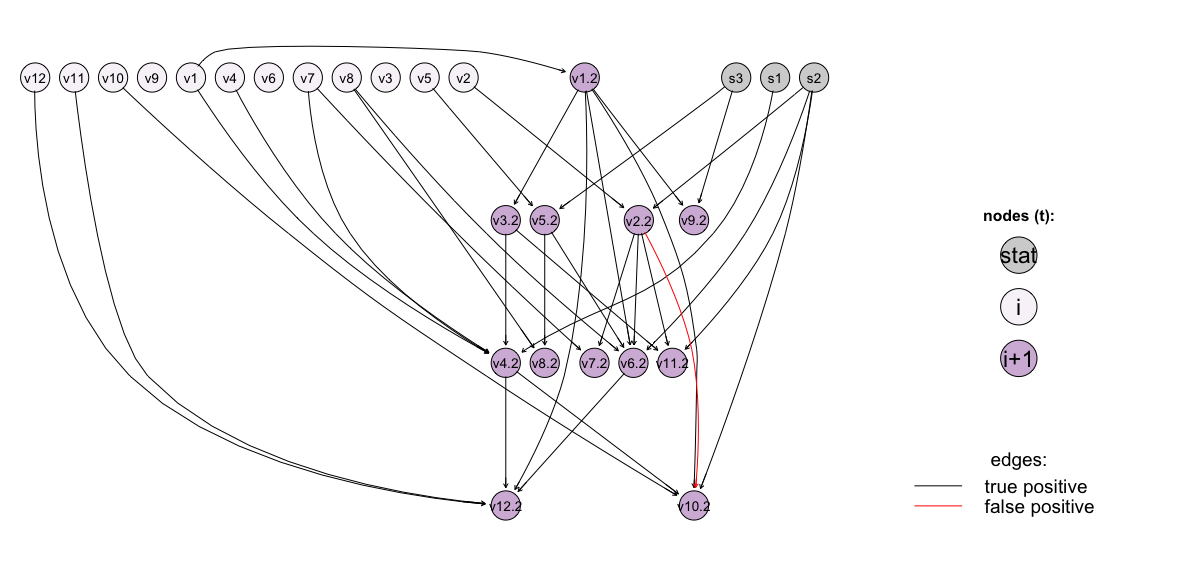}
\caption{Transition structure of MAP estimate of a 15-node DBN. Edge colors highlight differences/similarities between the maximum scoring structure found by iterative MCMC scheme and the ground truth transition structure. }
\label{fig:DBNcomp}
\end{figure}

Here we consider a simulated example of a DBN consisting of 12 dynamic and 3 static nodes. The data includes samples generated from five consecutive time slices. The syntax of structure learning functions then is the same as for usual Bayesian networks. Here we again learn the MAP estimate and optimize the search space with the function \code{\imc}. 

\begin{CodeChunk}
\begin{CodeInput}
R> DBNscore <- scoreparameters("bge", DBNdata,  DBN = TRUE, 
+ dbnpar = list(samestruct = TRUE, slices = 5, b = 3))

R> DBNfit <- iterativeMCMC(DBNscore)

R> plotdiffs.DBN(DBNfit$DAG, DBMmat, struct = "trans", n.dynamic = 12, 
+ n.static = 3)
\end{CodeInput}
\end{CodeChunk}

Figure \ref{fig:DBNcomp} demonstrates the transition structure of MAP estimate of a DBN learned from \code{DBNdata}. The MAP transition structure found by \code{iterativeMCMC} is very close to the ground truth structure with just one false-positive edge. Typically for DBNs, we observe many transitional edges connecting the same variable in neighboring time points $i$ and $i+1$. 

%% file: chapters/Application.tex
\section{Applications}
\label{sec:app}

\cite{Kuipers2018} used \pkg{BiDAG} for learning structures of Bayesian networks that characterize mutation profiles across cancer types and novel subtypes. The dataset included mutational profiles of $N=8198$ tumor samples across 22 cancer types. For $n=201$ significantly mutated genes, a Bayesian network-based clustering approach was used to define clusters of tumor samples, such that a Bayesian network represented each cluster center. Structure learning was performed in two steps. In the first step, the function \code{iterativeMCMC} from \pkg{BiDAG} was used to optimize the search space. In the second step, sampling was performed with \code{partitionMCMC} on the optimized search space. Posterior model selection was performed based on the sample of 100 DAGs from the posterior distribution with a posterior threshold of 0.5. The code for unsupervised clustering as well as the example of using \pkg{BiDAG} for Bayesian network-based clustering can be found at https://github.com/cbg-ethz/pancancer-clustering. \par
\cite{Kuipers2018} discovered networks both in supervised and unsupervised settings. As a demonstration here we show how \pkg{BiDAG} can be used to characterize cancer subtypes in a supervised setting and follow the learning steps described in \cite{Kuipers2018}. 
 
\begin{figure}[b!]
\centering
\includegraphics[width=\textwidth]{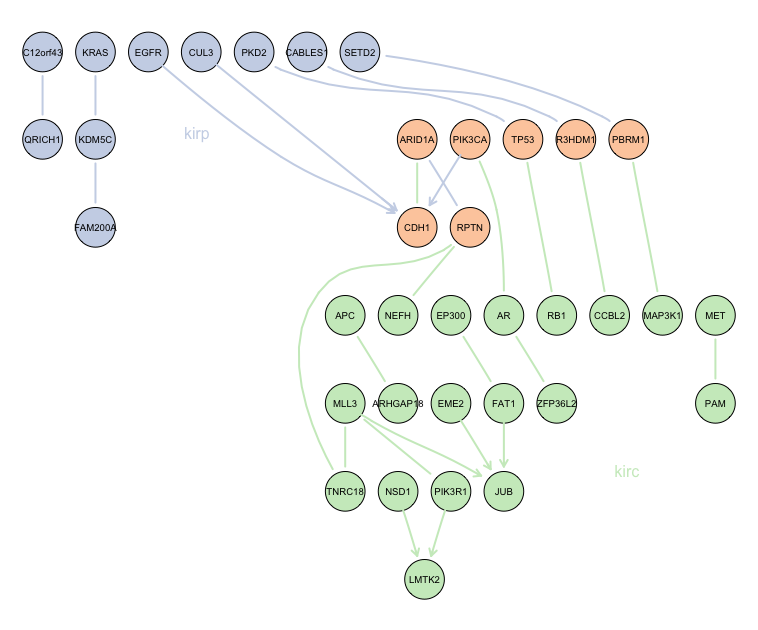}
\caption{Joint graph representing KIRC and KIRP MAP CPDAGs obtained by \code{iterativeMCMC}. Only nodes that have at least one connection are shown. Blue and green nodes/edges are specific to KIRP and KIRC graphs, respectively. Orange nodes have connections in both graphs.}
\label{fig:kirckirpMAP}
\end{figure}

 We analyze non-silent mutation data from two cohorts from The Cancer Genome Atlas (TCGA, \href{https://www.cancer.gov/tcga}{https://www.cancer.gov/tcga}). The cohorts represent two kidney cancer subtypes: renal papillary cell carcinoma (KIRP) and kidney renal clear cell carcinoma (KIRC). We include the most significantly mutated genes (q < 0.1) from both cohorts. Mutation data and corresponding lists of significantly mutated genes were obtained from \cite{https://doi.org/10.7908/c19c6wtf} and \cite{https://doi.org/10.7908/c10864rm}. Additionally, we have included connected genes from KIRP and KIRC networks discovered by \cite{Kuipers2018}. Both pre-processed datasets are accessible in the \code{BiDAG} package.
 
 Following the methods in \cite{Kuipers2018}, we use a prior derived from the protein-protein interaction database STRING \citep{STRING}. The edges that are not among interactions in the STRING database are penalized by a factor of 2 for graph inference. The database is being constantly updated, and known interactions between genes have changed considerably since the analysis reported by \cite{Kuipers2018} was performed. Here, we use the most recent version 11.0 of the database. In \pkg{BiDAG}, the function \code{string2mat} transforms the downloaded list of interactions from STRING into a matrix, which can be used for blacklisting or penalizing of edges in \code{BiDAG}.

We run \code{iterativeMCMC} to find MAP DAGs representing the KIRC and KIRP subtypes and corresponding equivalence classes. Figure \ref{fig:kirckirpMAP}, produced by the function \code{plot2in1}, shows edges from both discovered CPDAGs in one graph. The genes \textit{TP53, PIK3CA, ARID1A, PBRM1, CDH1, RPTN, R3HDM1} are connected to other nodes in both subgraphs representing KIRP and KIRC cohorts. However, the  subgraphs  do not share any edges.\par

\begin{figure}[b!]
     \centering
     \begin{subfigure}[b]{0.44\textwidth}
         \centering
         \includegraphics[width=\textwidth]{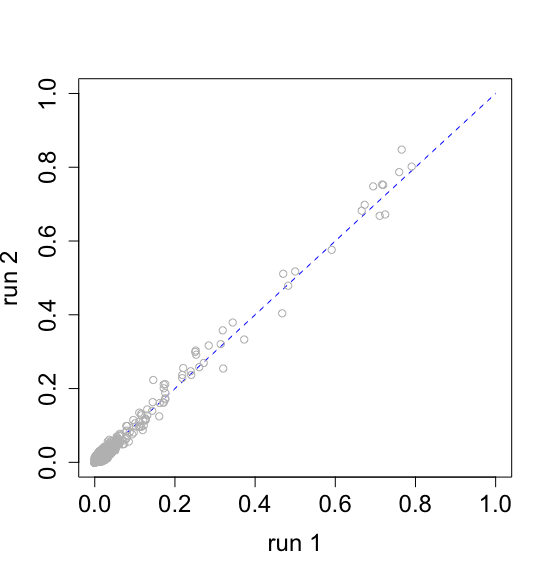}
         \caption{KIRP cohort}
         \label{fig:KIRP60}
          \end{subfigure}
         \hfill
     \begin{subfigure}[b]{0.44\textwidth}
         \centering
         \includegraphics[width=\textwidth]{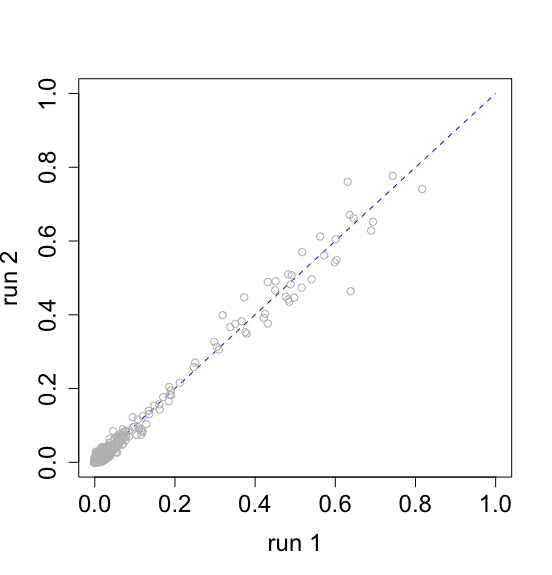}
         \caption{KIRC cohort}
         \label{fig:KIRC60}
     \end{subfigure}
   \caption{Convergence diagnostic plot: concordance of posterior probabilities estimates of single edges between pairs of \code{partitionMCMC} runs.}
\label{fig:concKK}
\end{figure}

We proceed with \code{partitionMCMC} to understand how confident we can be about the discovered mutational interactions. To check convergence we use the \code{edgep} and \code{plotpcor} functions as in the previous section. The result is shown in Figure \ref{fig:concKK}. There are no visible convergence problems, and all the points are close to the diagonal.

 \begin{figure}[htb!]
     \centering
     \begin{subfigure}[b]{0.45\textwidth}
         \centering
         \includegraphics[width=\textwidth]{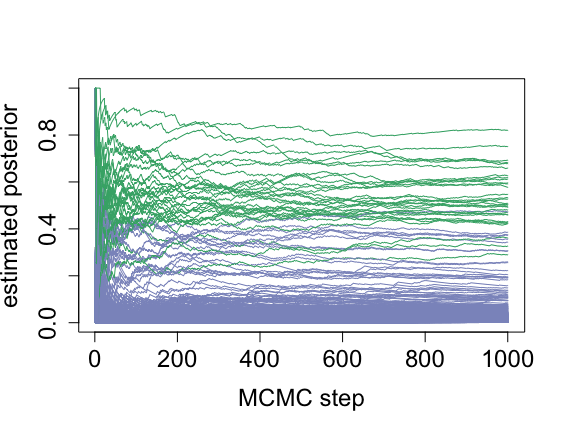}
         \caption{KIRC, function \code{plotpedges}}
         \label{fig:KIRCpedges}
     \end{subfigure}
     \hfill
     \begin{subfigure}[b]{0.45\textwidth}
         \centering
         \includegraphics[width=\textwidth]{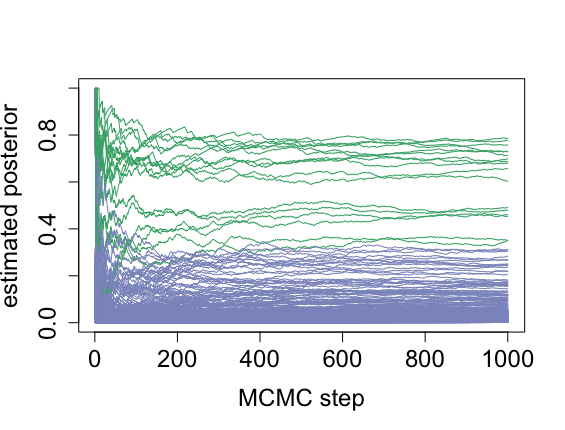}
         \caption{KIRP, function \code{plotpedges}}
         \label{fig:KIRPpedges}
     \end{subfigure}
   \caption{Convergence of posterior probabilities of single edges: a pair of MCMC chains were run for each of KIRP and KIRC datasets. Posterior probability traces were obtained by applying Equation \ref{postedges} ($m=1$) at each MCMC step ($M=1,2...,1001$). Green lines correspond to traces of posterior probabilities of the edges of the estimated MAP CPDAG.}
\label{fig:pedgestcga}
\end{figure}

\begin{figure}[b!]
\centering
\includegraphics{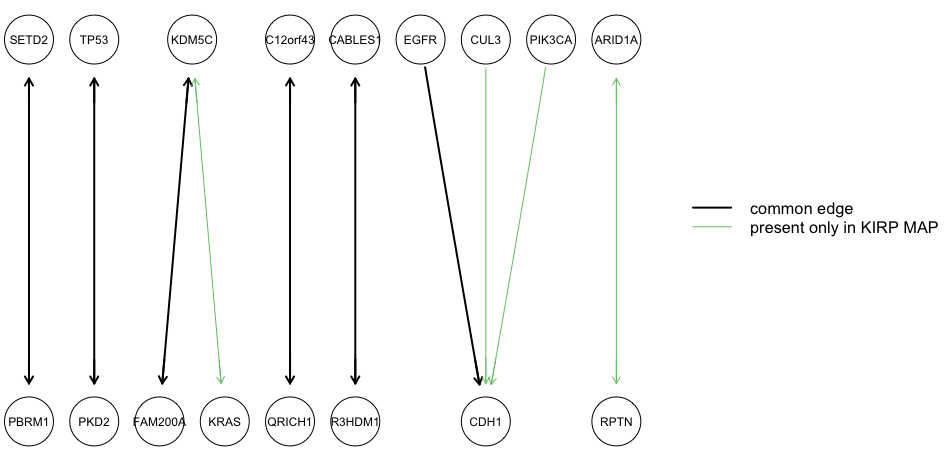}
\caption{Comparison between MAP CPDAG and consensus PDAG learned from the \code{kirp} dataset. MAP CPDAG was obtained with \code{iterativeMCMC}. Consensus PDAG was obtained by averaging over a sample of DAGs converted to CPDAGs and keeping the edges whose posterior is bigger than 0.5.}
\label{fig:KIRPgraph}
\end{figure}
Another useful plot for checking convergence depicts how the posterior probabilities of single edges change through MCMC iterations. In Figure \ref{fig:pedgestcga} we can see that posteriors of the vast majority of edges stabilize in both cases after a short burn-in period. Edges of the MAP CPDAG (highlighted in green) reach higher posterior probabilities than almost all other edges. However, many edges of MAP structures converge to posterior levels below 0.5. The posterior probability of an edge can be interpreted as a measure of confidence in the edge based on the data. \cite{plus1} have shown that when the data is noisy or scarce, the MAP graph may include a lot of edges with low posterior probability, many of which turn out to be false positives. We perform model averaging based on a sample of DAGs obtained by partition MCMC and remove the edges with a posterior less than 0.5. As we have seen in Section \ref{examples}, this approach can help to remove false-positive edges, while keeping most of the true positives.

 Figure \ref{fig:KIRPgraph} depicts differences between MAP and consensus graphs for the KIRP cohort. Six edges out of ten passed the posterior threshold of 0.5. In the KIRC cohort, similarly, 11 out of 18 edges of MAP CPDAG passed the threshold of 0.5. 
 
\begin{figure}[t!]
\centering
\includegraphics{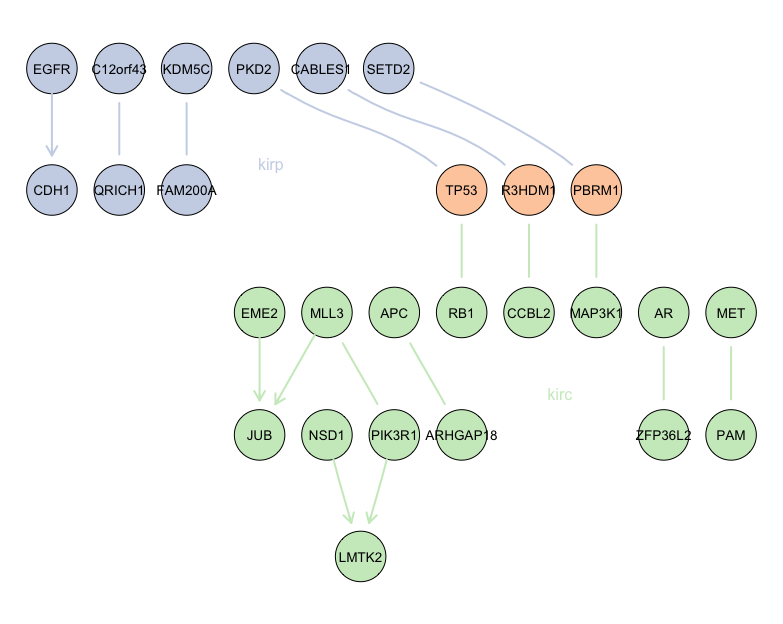}
\caption{Joint graph for KIRC and KIRP  cohorts obtained averaging over a sample from posterior distribution obtained by \code{partitionMCMC}. Only nodes that have at least one connection and whose posterior probability is higher than 0.5 are shown. Blue and green nodes/edges are specific to KIRP and KIRC. Orange nodes have connections in both graphs.}
\label{fig:KIRCKIRPsamp}
\end{figure}

Figure \ref{fig:KIRCKIRPsamp} visualizes the consensus models for KIRP and KIRC in one graph. Many of the discovered edges correspond to those found previously by \cite{Kuipers2018} for the respective cancer subtypes. Although we have followed the analysis steps from \cite{Kuipers2018} on the same set of tumors, we used a different prior and a different set of genes. Consequently, the discovered networks display some differences.  For example, we discovered an edge between \textit{CCBL2} and \textit{R3HDM1}, but the gene \textit{CCLB2} was not included in the list of genes analyzed by \cite{Kuipers2018}.

%% file: chapters/Runtime.tex
\section{Runtime}
\label{runtime}
The number of MCMC iterations used in the order MCMC scheme by default is $6n^{2}\log{n}$, for a network with $n$ nodes. Each MCMC iteration requires the computation of the score $R(\prec^{'}\mid D)$ of at least one proposed order. When implemented naively, the complexity of scoring an order is exponential, $O(n^{K+1})$, where $K$ is the maximum number of parents allowed in the scheme. This brings the total chain complexity to $O(n^K n^{2} \log{n})$. For efficient implementation, we use the approach and computational optimizations described by \cite{plus1} and pre-compute the quantities needed to score an order at each iteration of the MCMC scheme. We refer to this step further as pre-computing the score tables. This reduces the complexity of the chain by a factor of $n^K$  to $O(n^{2} \log{n})$ \citep{plus1}. Of course, the complexity of computing the score tables remains exponential, but now it is independent of the number of MCMC iterations as it has to be done only once. In addition, using the search space $\mathcal{H}$ instead of restricting the number of parents to a hard limit $K$ reduces the complexity of computing the score tables to $O(n K^2 2^K)$ or $O(n K^3 2^K)$ depending on the score type. As a result, computing the score tables is more efficient for all $n$ and $K$ when compared to the naive implementation. \par 

When the score tables are pre-computed, the complexity of the MCMC scheme is polynomial in the size of the number of nodes of the network, $n$, such that the algorithm is applicable to large networks with hundreds of nodes. The computation of the score tables is exponential in the maximal parent set size $K$, so $K$ imposes a feasibility limit on the implemented algorithms. While no hard limit for $K$ is required in \pkg{BiDAG}, for large $K$, the score table computations can become prohibitive.\par

\begin{figure}[b!]
\centering
\includegraphics[width=100mm,scale=1]{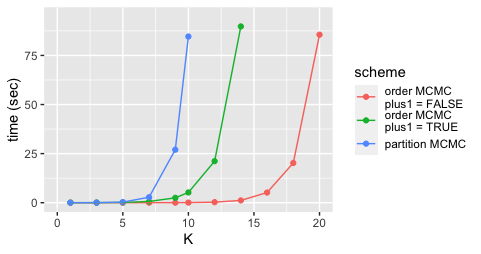}
\caption{Time needed to compute a score table for a node with $K$ parents for a network with $n=100$ nodes.}
\label{fig:timeK}
\end{figure}
\par

Figure \ref{fig:timeK} shows how much time is needed to compute the score tables for a node with $K$ parents for a network with $n=100$ nodes. All timing measurements were carried out on a 2.3 GHz Intel Core i5 processor. For $K>7$, the differences in runtimes of different MCMC schemes become substantial. As expected, the runtime is lowest for order MCMC sampling in $\mathcal{H}$ and highest for partition MCMC. Building score tables in the core search space $\mathcal{H}$ (\code{plus1}=FALSE) requires shorter time than in its extended version $\mathcal{H}^{+}$ (\code{plus1}=TRUE), but scoring a node with up to 14 parents is feasible in both cases for the order MCMC scheme. Note, however, that most real-world networks are much sparser than that. For example, in 30 networks found in the BN repository  (\href{http://www.bnlearn.com/bnrepository/}{http://www.bnlearn.com/bnrepository/}) the average parent set size is 1.4, while the maximum parent set size is 13. \par

The parameter \code{hardlimit} of structure learning functions \code{orderMCMC} and \code{partitionMCMC} ensures that the search space contains only nodes with parent set sizes not exceeding this limit. As mentioned in Section \ref{sec:package}, \code{iterativeMCMC} stops extending a node's parent set when the \code{hardlimit} has been hit for this node, but it can still expand the parent sets of other nodes until they all reach the limit or the score does not improve further.\par

%% file: chapters/Discussion.tex
\section{Discussion}

The \proglang{R} package \pkg{BiDAG} implements flexible MCMC schemes for structure learning and sampling of Bayesian networks.  The iterative MCMC scheme can be used to search for a MAP graph and to optimize the search space, while partition and order MCMC can be used for sampling from the posterior distribution. Order MCMC converges faster and is computationally less demanding than partition MCMC, but only the latter provides an unbiased sample of the posterior. Other tools for structure learning either focus on finding one best solution or implement Bayesian approaches, which are not feasible for large networks due to computational costs or slow convergence. \pkg{BiDAG} is the first package available for efficient sampling of DAGs with hundreds of nodes. In the future, we plan to implement features that could potentially reduce the runtimes of partition and iterative MCMC schemes. In the iterative MCMC scheme, we consider adding algorithms other than PC for defining the initial search space.  Furthermore, the convergence of partition MCMC could be improved with the addition of new moves.